\documentclass[12pt]{article}
\usepackage{amssymb}

\begin{document}
\begin{center}
{\Large\bf A Categorical Framework for\\[0.2cm]
Quantum Theory}\\[0.5cm]
{\bf Thomas Filk${}^1$, Albrecht von M\"uller${}^2$} \\[0.2cm]
{${}^{1,2}$Parmenides Center for the Study of Thinking, Munich}\\
{${}^{1}$Institute for Physics, University of
  Freiburg,}\\
{Hermann-Herder-Str.~3, D--79104 Freiburg} \\
{${}^2$ Institute for Philosophy, University of Munich\\
SISSA, Via Beirut 9, I--34151 Trieste}
\end{center}

\thispagestyle{empty}


\vskip 1cm

\begin{abstract}
Underlying any theory of physics is a layer of conceptual
frames. They connect the mathematical structures used in
theoretical models with physical phenomena, but they also
constitute our fundamental assumptions about reality.
Many of the discrepancies between quantum physics 
and classical physics (including Maxwell's electrodynamics 
and relativity) can be traced back to these categorical 
foundations. We argue that classical physics corresponds 
to the factual aspects of reality and requires a 
categorical framework which consists
of four interdependent components: boolean logic, the 
linear-sequential notion of time, the principle of sufficient 
reason, and the dichotomy between observer and observed. 
None of these can be dropped without affecting the others.

However, in quantum theory the reduction postulate also 
addresses the ``status nascendi''
of facts, i.e., their coming into being. Therefore, 
quantum phyics requires a different conceptual 
framework which will be elaborated in this article. It is 
shown that many of its components are already present 
in the standard formalisms of quantum 
physics, but in most cases they are highlighted not so much
from a conceptual perspective but more from
their mathematical structures. The categorical frame 
underlying quantum physics includes a 
profoundly different notion of time which 
encompasses a crucial role for the present. 
\end{abstract}
\newpage

\section{Introduction}

Even today, many scientists and philosophers of science
still struggle with the fundamental concepts of quantum 
theory. It is not the mathematical formalism which is at 
the center of this struggle - apart from technical details 
the formalism is relatively simple - but it is more the
conceptual questions and the physical intuition related
to this mathematical framework which is unsatisfactory.
It has often been argued that our ways of thinking
rely on our every-day experiences, and, therefore, may
not fit to the realm of microscopic systems. But on
the other hand, we have far less problems dealing with 
geometry in 26 dimensions or with the topology of 
infinite dimensional spaces. It is not so much a lack of
imagination and visualization which is at the heart
of the problems, but rather an apparent contradiction 
of certain aspects of quantum theory with our most
basic assumptions about reality. 

What are some of these aspects of quantum theory which
scientists struggle with? Surely, one aspect is related to
the intrinsic indeterminacy of quantum theory. We are
used to indeterminacy in every-day situations, but 
whenever we refer to probability theory, we consider
this indeterminacy as a lack of
knowledge. In some way or another we are reluctant to
give up Leibniz' principle of sufficient reason which
states that ``nothing takes place without a sufficient 
reason; in other words, that nothing occurs for which 
it would be impossible for someone who has enough 
knowledge of things to give a reason adequate to 
determine why the thing is as it is and not 
otherwise.''\cite{Leibniz}
This principle seems to be violated in quantum
theory, at least in those interpretations which include 
the reduction postulate. The violation of Bell's
inequalities \cite{Bell} in quantum physics \cite{Aspect}
seems to indicate that in some cases even the 
mere assumption of a predeterminded cause
for the behavior of quantum systems (e.g., the
observed deviation of an atom in a Stern-Gerlach 
set-up) leads to contradictions. 

Another conundrum of quantum theory is related to
its non-locality, in particular in the context of
EPR-scenarios \cite{EPR}, i.e., in cases which involve
entangled states. Quantum states can be
highly non-local, but local measurements 
can lead to an instantaneous reduction of the
whole state. Two fundamental concepts seem to
contradict each other: the locality principle 
of relativity and the reduction postulate of
quantum theory. In addition to its non-local
aspects, entanglement expresses a type of relationship
which is unknown in classical physics. The
concept of facticity is put into question
when entangled states are involved. Rovelli
in his relational interpretation of quantum 
theory describes this situation as: 
``... a physical fact, its being true, or
not true, must be understood as relative
to an observer...'' \cite{Rovelli}. 

Still another weird aspect of quantum theory 
are ``superpositions''. Even though superpositions
only refer to vector representations of
quantum states, there is a more general phenomenon
behind it: In quantum theory there are no states
which are non-dispersive with respect to all
observables. This statement is trivial for mixed
states, but it also refers to pure states. 
In other words, for any state there exist
observables such that repeated measurements 
of the same observable on systems prepared in this 
same state do not always yield the same results. 
Again, the formalism does not allow to attribute 
this fact to an imperfect preparation of the state 
(which is not separated with respect to certain ``hidden''
variables), but it is something intrinsic to
quantum states.

As a final aspect we mention the lack of a
rigid spatiotemporal ordering of events. 
Events do not happen at a particular location in space 
and at a particular moment in time. The spatial 
``smearing out'' of particles 
as expressed by wave functions is well known
and partially related to the non-boolean aspect
of quantum theory: a particle is not ``either'' in
a certain volume element ``or'' not, but according
to the extension of the wave function it can be
to a certain degree inside a volume element ``as well
as'' outside this element. Usually less emphasized
is the temporal aspect of this phenomenon: a lack of 
sequentiality of events. This partial loss of a
sequential time will be one of the major subjects
of this article.

Various interpretations of quantum theory
have addressed the problems mentioned above.
Physicists favouring a more formal or
positivistic approach may deny any conceptual
problems of quantum theory because the 
mathematical concepts are well
defined and free of contradictions, and the
associations of mathematical structures with
physical observables are equally well defined 
and in full agreement with all experiments. 
Protagonists of the many-world interpretation
\cite{Everett,DeWitt}
may deny a problem with the non-locality of 
the reduction process or with indeterminacy,
because in the many-world interpretation there 
is no reduction.
Supporters of a subjective or an information based
interpretation of quantum states see no
problem in the reduction process, because 
according to their understanding, reduction 
is merely a change of knowledge about a system. 
Still, for many physicists the alternatives offered 
by these interpretations are unacceptable for
reasons of their own. 

Two major assumptions about
quantum theory will be made in this article: (1)
the quantum state of a system is related to an
ontic element of reality (thus excluding purely
information based and subjective interpretations
of a quantum state), and (2) the reduction
process too is related to an ontic element of
reality (hence, we will mention but not focus
onto the implications which 
the many-world interpretation has for the 
applicability of our approach). Furthermore, we will
also not discuss explicit alterations of the
standard formalism of quantum theory, like the
the approaches of Karolyhazy \cite{Karolyhazy} and 
Penrose \cite{Penrose}, which attribute the reduction
process to an influence of gravity, or the GRW
formalism \cite{GRW} and relativistic extensions
\cite{Tumulka}, which add to the Schr\"odinger 
equation a non-linear stochastic term. In these
cases the reduction of the wave function is attributed 
to reduction centers. We will also
not discuss interpretations which invole
a causation into the past (like, e.g., in \cite{Cramer})
or which are ``superdeterministic'' by including the
future decisions of experimenters about which experiment
to make into the initial conditions of quantum systems
(like, e.g., the approach of Palmer \cite{Palmer}).   
The reason why we will not elaborate our conceptual
framework in the context of these approaches is not
so much that our framework is not applicable, but
rather that it would require a refinement and
adaptation for each single case which would go far
beyond the scope of a single article.

Our approach to the counterintuitive
aspects of quantum theory is less formal and more
conceptual. First, we want to analyse, 
which ``hidden assumptions'' we often make when we 
try to interprete quantum mechanical issues
and why this hidden assumptions in the context of
quantum theory seem to lead to
contradictions with our understanding of how 
reality ``should be''. These hidden assumptions
are no surprises: We are aware that, e.g.,
we tend to apply causality in the sense of Leibniz'
principle of sufficient reason, or that we
apply boolean (``either-or'') logic to physical 
attributes like location, energy, and momentum,
or that we like to put ourselves into an external
perspective (in contrast to the ``participatory
universe'' of Wheeler\cite{Wheeler1}). We apply 
these assumptions
successfully in the context of classical physics,
and we know that they lead to contradictions with 
quantum theory, but we often do not see the conceptual
alternatives. Therefore, as a second step,
we will suggest a replacement of these concepts 
which work so well in the context of
classical physics. This replacement will
not be on the formal mathematical level - such
concepts exist, e.g., the axiomatic algebraic
approach \cite{Haag} or the notion of non-boolean
lattices \cite{Giuntini} - but rather on the
level of conceptual thinking. Therefore, our approach
does not aim at a replacement of the existing 
mathematical formalism, but instead it elaborates 
the underlying categorical assumptions
thus allowing for a coherent interpretation.
The purpose of this article could in a way
be summarized as: Making quantum physics thinkable.

One might ask whether such a non-mathematical 
conceptual foundation is necessary or what the
advantages might be of having such a categorical 
framework at hand. We will address this question
in the next section (Sec.\ \ref{sec_Why}) 
but a brief answer will be that the difficulties
we have in formulating a consistent and
satisfactory unification of general relativity and
quantum theory might indicate that some of the basic
notions we use in the context of these two theories
are incompatible and not yet 
fully understood. For this reason it not only seems
legitimate but necessary to rethink the 
conceptual underpinnings in these theories.

In Sec.\ \ref{sec_becoming} we will address the 
issue that quantum theory mainly deals with the
``coming into being'' of reality, while classical
Newtonian physics is more about the factual
aspect of reality. In our opinion, this is the
major difference between classical physics and
quantum theory, and it is this difference 
which makes a richer conceptual
framework necessary for quantum theory. In Sec.\
\ref{sec_Categories} we introduce the notions
of a category, a categorical slot, a categorical 
apparatus and 
a categorical framework in the more philosophical
meaning which we will use in this article.
After these preliminary notions we will describe in 
Sec.\ \ref{sec_F} the categorical assumptions we 
make in the context
of classical, factual physics. In Sec.\ \ref{sec_E}
we will introduce the categorical apparatus 
necessary to address questions related to
``reality in the making'', i.e.\ the categorical
apparatus of quantum theory. Both apparatus taken together,
the categorical apparatus related to factual classical
physics and the categorical apparatus related to quantum
theory, make up the whole categorical framework 
necessary to address reality as such.

One category in the apparatus related to quantum theory
- the non-sequential ``time-space of the present'' -
is of particular importance. We will argue that
certain ingredients in the formalism of quantum theory
indicate that for a full understanding we may need a
theory of the present. In contrast to
the other categories which are well-known from the
standard formulation of quantum theory, this aspect 
is usually less emphasized in the standard
presentations of quantum theory.
Therefore, we will devote Sec.\ \ref{sec_Time}
to this concept. A brief summary as
well as an outlook to other applications of our
scheme will conclude this article. Because of the
unfamiliar approach taken in this article, 
we will describe some of the physical
aspects and their relation to the philosophical
concepts in more detail than would otherwise
be necessary.  

\section{Why a categorical framework?}
\label{sec_Why}

Before we introduce the categorical framework 
which in our opinition is most appropriate for
dealing with classical and quantum physics, 
and ``reality'' in general, 
we want to justify why it might be necessary
to develope such categories or concepts and
to clarify the relations among these concepts. 
Why is it not sufficient to
refer exclusively to the mathematical formalism 
which allows to make predictions for the outcomes 
of future experiments? 

A famous answer to this question is known from
Einstein: {\em It has often been said, and certainly 
not without justification, that the man of science is a 
poor philosopher. Why then should it not be the right 
thing for the physicist to let the philosopher do the 
philosophizing? Such might indeed be the right thing 
to do at a time when the physicist believes he has at 
his disposal a rigid system of fundamental concepts 
and fundamental laws which are so well established 
that waves of doubt can't reach them; but it cannot 
be right at a time when the very foundations of physics 
itself have become problematic as they are now. At a time 
like the present, when experience forces us to seek a newer 
and more solid foundation, the physicist cannot simply 
surrender to the philosopher the critical contemplation 
of theoretical foundations; for he himself knows best 
and feels more surely where the shoe pinches. In looking 
for a new foundation, he must try to make clear in his 
own mind just how far the concepts which he uses are 
justified, and are necessities.}\cite{Einstein}

A present day example for this necessity can be found
in the efforts to find a conceptually satisfying 
unification
of general relativity and quantum theory. One
approach to such a unification is to apply 
and extend the existing mathematical formalism 
to new areas and to hope that the results are still
meaningful. This is done in the context of
canonical quantization, loop quantum gravity, 
Wheeler-DeWitt type of approaches or even 
string theory. Another direction to go is
to reconsider the fundamental concepts - space, time, 
matter, reality etc. - of the existing theories and 
to clarify the exact meaning in which
these concepts are used, why these attributed meanings 
may be incompatible with each other, and by what
concepts we might sensibly replace them.
An attempt of this kind is made in this article.

Another reason to analyse the fundamental concepts 
from a more general perspective and, in particular, 
to relate them to ideas which are not from 
mathematical textbooks (later we will use 
expressions like ``autogenesis'', ``parataxis'', 
``self-referentiality'', etc.) is that these 
concepts actually do occur outside the realm of
quantum physics, for instance in the areas of
consciousness and selfhood or of literature
and art. In these areas we are used to expressions which 
often refer to a vague feeling or an impression and 
which lack mathematical rigor. Nevertheless, we 
often use these concepts with considerable success 
in the development of ideas and the establishment 
of relations which we would miss without
these concepts. In quantum theory we are
in the lucky situation, that we have a rigorous
mathematical formalism which we can use in making
exact statements or predictions, and which we can
also use in making the vague concepts mentioned above
more rigorous. What we often lack
is the intuitive notion which helps us to develop 
new ideas. Therefore, our aim is on the one hand
to keep the rigorous mathematical formalism for
deducing precise conclusions, and, on the other hand,
to enlarge this tool by a more intuitive but, 
nevertheless, equally rich framework of categories
which, when being aware of them, can be used to
get a deeper understanding of quantum physics
and, more general, of reality.  

\section{Quantum physics - 
the ``status nascendi'' of reality}
\label{sec_becoming}

Why does quantum theory need a different set of 
categories as compared to the classical physical 
theories like Newtonian mechanics, Maxwells's field 
theory, or even special and general relativity? Why 
are already the basic questions addressed by
quantum theory of a fundamentally different nature? 
We argue, and this will be elaborated in more detail
in this section, that quantum theory is a theory
about how facts come into being, how facts ``are born'',
i.e.\ the ``status nascendi'' of reality. In contrast,
classical physics deals with facts, not the
coming into being of facts. This fundamental 
difference requires the different categories 
which we will introduce in the next sections.

In classical physics we mainly talk about facts. 
By facts we usually mean configurations or
constellations of objects which we see 
(or otherwise detect). From these configurations
we can deduce that certain events have happened.
Even if we talk about ``events'' in the context of
classical physics, we usually refer to the facts
which evolved from these events. 
In this sense we will use the word ``fact'' to denote
the traces which events have left behind in the 
constellation
of our present reality. An obvious example for such a
trace is a fotograph which shows us a picture of
a situation in the past. Similar examples are fossiles
or written documents. Less obvious but of a similar kind
are the traces left behind in our brain which allow us
to remember certain situations. Another example are
scattered particles whose momentum or energy carry
information about the scattering event. Only what 
leaves factual traces qualifies for the notion of a 
``real event''. 

Even in cases where we seem to observe a certain 
event, what we perceive and observe are already traces: 
the photons which have been scattered by objects which 
took part in an event or the sound waves which have been
emitted by a clash of certain objects. Even when we
refer to the event of observation itself, by the moment 
we become aware of this event 
we are already talking about facts. The closest we come
to taking part in an event itself is the phenomenon of
becoming aware of something. However, now the problem of 
consciuosness enters and we are leaving the realm of
known physics. (In \cite{FM} we elaborated on the
similarity between the conceptual frameworks of quantum
theory and consciousness.)

The deterministic nature of classical physics
leaves no room for possibilities or alternatives 
apart from the epistemic uncertainty which arises 
because of our lack of knowledge about certain 
situations. In principle,
the present state already determines the future
states in a similar way as from the present state
one can deduce the past. In the realm of classical
physics it is the
second law of thermodynamics (the increase of
entropy in non-equilibrium systems) which 
makes it easier for us to deduce events in the
past from facts in the present as compared to
predicting events in the future from the
present situation. 

In contrast, quantum theory refers to the transition
from possibilities (or potentialities) to facts. A wave 
function, or, more general, a quantum state, carries 
information about the past of the system, but with 
respect to
possible events in the future it only allows to make
probabilistic predictions. In quantum theory, this
probabilistic nature is intrinsic, not a lack
of knowledge about the present state. The transition
from this ``sum of possibilities'' to a single
factual result is what we define as the ``event'',
and in this sense quantum theory addresses the
coming into being - or the ``status nascendi'' - 
of facts.  

At this stage some remarks about our notion of ``event''
are in order. Whenever we are referring to non-quantum physics,
the expression ``event'' is used in the same sense as
it is used, e.g., in the context of special or general
relativity: Events are the nodes of the gist of space-time.
In most cases, they are idealized to be ``point-like''. 
Events can be a conglumerat of more elementary
events and in many cases we even assume that there are
``basic'' events which cannot be separated into subevents
anymore.

When we talk about events in the context of quantum
theory we should, strictly speaking, distinguish two
types of events. One type of event refers to the transition
from possibilities to facts in the reduction process.
When there is a danger of confusion, we will use the
expression ``transgressive event'', emphasizing
the transitional aspect of ``becoming a fact''. An attempt to
give a more precise meaning to this type of event can be
found in \cite{Haag2}. A second type of event will be what 
in discussions about quantum theory sometimes is called 
``virtual event''. These type of events often refer to an 
interpretation of aspects of the formalism (e.g., the terms in a
perturbative expansion, like the emission of a photon
by an electron in a scattering process). They do not
lead to facts individually, but only as parts of a
transgressive event.  

\section{A categorical framework}
\label{sec_Categories}

In this section we will introduce the notion of a catgeorical
apparatus and a categorical framework. Loosely speaking, a
categorical apparatus consists of a set of categories which
are mutually interdependent, and a categorical framework 
consists of two (or more) categorical apparatus which, 
taken as a whole, allow to formulate any meaningful statement 
with respect to the realm addressed by the categories. 
The notions of ``category'' and ``categorical'' are not 
used in a mathematical but a philosophical sense.   

The notion of a category has a long history in philosophy, 
and different thinkers have often attributed different meanings
to this notion. Fortunately, we will not need a rigorous 
definition but only the general idea (actually, in the
context of the approach developed here, a comprehensively
well-defined notion of ``category'' may not even be possible). 
{\em Categories} are the most fundamental interface between
reality and cognition. They constitute primordial 
assumptions about the nature of phenomena. For instance, 
the assumption that all physical processes are 
deterministic and can be described by an equation 
of motion is a categorical assumption 
which holds in Newtonian physics but not in quantum physics.
As we shall see, however, categories do not come as 
isolated entities, but they are mutually interrelated. 
For such a set of interdependent categories we introduce 
the notion of a {\em categorical apparatus}.

The general structure of a categorical apparatus
consists of various slots that are to be filled
with specific, interrelated categories.  
This means that categories which we insert into
a certain slot have consequences with respect to the 
other categories. These interdependencies
of categories are so rigid, that one category
predetermines the other categories, and any attempt to 
substitute a given category by a different assumption 
leads to inconsistencies with the other categories. 
The slots within a categorical apparatus will be called
a {\em categorical template}.

To make the idea more crisp let us give an example which
we will study in more detail in Sect.\ \ref{sec_F}:
the categorical apparatus of classical (non-quantum) physics,
by which we mean Newtonian mechanics, Maxwell's field theory 
and relativity. We argue that the following four slots make 
up a categorical template (for a more philosophical account
of this structure see \cite{AvM,AvMa,AvMb}):
\begin{enumerate}
\item
What is the relation between physical events? 
\item
How can we combine predications about physical observables?
\item
What is the relation between the observer and the observed?
\item
Which ordering structures are attributed to space and time? 
\end{enumerate}
Underlying classical physics are the following four
categorical assumptions:
\begin{enumerate}
\item
The events which make up physical processes 
are causally deterministic.
\item
For all physical attributes
the {\em tertium on datur} holds, i.e., in a given 
context only one of the predicates, ``$a$'' or 
``not $a$'', can be true.
\item
The observer can be considered as separated 
from the observed and the acquisition of 
information is possible without 
influencing the observed system.
\item
At each instant in time, the location of physical
bodies is determined by their relative distances.
Along any world-line, events are sequentially ordered.
\end{enumerate}
It seems obvious that if we
change one of the categories, e.g., the sequentiality
of time (``for any two events $a$ and $b$ along a 
world-line either $a<b$ or $b<a$ holds'') this has
a huge impact on the other categories. E.g., the 
standard notions
of causality or determinism can no longer hold if
this assumption is dropped.

We claim that all physical theories imply certain 
underlying categorical assumptions. This refers 
only partially to the mathematical formalism per se 
but more to its association
with physical phenomena. It will turn out, however, 
that not all questions related to reality can be
formulated meaningfully within only one categorical
apparatus, e.g., the frame defined by the classical 
categorical apparatus given above. Quantum physics
requires a different categorical apparatus. None of
the four classical categories holds in quantum physics, 
i.e., all categories have to be changed. For this
reason, attemps which try to cope with quantum 
theory by relaxing only one of the categories
listed above are bound to fail. The categorical
apparatus given above will be applicable whenever we
make statements about the factual aspect of reality,
therefore we will refer to it as the {\em F-scheme}.
On the other hand, quantum theory also deals with
the ``coming into being'' of facts. Whenever
we address the ``statu nascendi'' aspect of 
reality, i.e., make ``event-related'' statements, the
second categorical apparatus, the so-called {\em E-scheme}
will be needed. Both apparatus together, the
F-scheme and the E-scheme, constitute the complete
{\em categorical framework} that allows to address 
physical reality in a comprehensive way. 

\section{The categorical apparatus of factual reality}
\label{sec_F}

As we have mentioned before,
facts are traces left behind by events. In
some extreme cases the trace-character is immediately
obvious (like in a fotograph, a fossile or the connectivities
in the brain leading to a memory). Other examples are
the click of an electron in a detector 
or the polarization of a photon which scattered
from some surface. In general, the traces of even
a single event become distributed over a huge number of 
degrees of freedom and may in practice never be observed 
in their totality.  But when we observe reality we mostly 
look at facts, and classical
physics is essentially a theory about facts. Even though 
the equations of motion allow to make predictions about
future states of a system, classical physics is not
concerned with the event of ``coming into being'' 
itself. Actually, in the framework of all variants
of classical physics (including relativity)
this problem cannot even be addressed. 
The deterministic character of the 
equations of motion lead in an almost automatic way 
to a block type universe in which everything which is,
was, or will be has the same degree of facticity. 

For addressing questions related to the factual aspects
of reality, we refer to the categorical apparatus
mentioned in Sec.\ \ref{sec_Categories}. The four slots -
(1) relation between physical events, (2) combinations of 
predications, (3) relation between observer and 
observed, (4) ordering
structure of time and space - are filled with the 
following categories: (1) processes are deterministic 
(causally closed), (2) a predicate about a physical 
attribute is either true or false, (3) an observation
has no influence on the observed, (4) space and time 
allow for an ordering structure: events along a world
line can be sequentially ordered and for objects in 
space there exists a metrical and topological ordering. 
In the following sections we shall discuss these 
categories in more detail.

\subsection{The deterministic nature of physical
processes}

In the context of classical physics, the changes in
state space are described by equations of motion. 
If the state of a closed physical system is given
at a certain instant $t$, the equations of motion determine
the state for any other instant $t'$. This is also true of
Maxwell's theory of electromagnetism and of special and
general relativity.\footnote{In particular with respect to 
general relativity we should mention, that certain initial
conditions may lead to singularities in the solutions which 
make a determination of this
solution of the equations beyond the singularity impossible 
or even meaningless. Such singularities indicate a break-down
of the classical theory and, thereby, of the classical
categorical apparatus or F-scheme. Effects of quantum gravity
may become relevant.} 

At this point we should emphasize that we are not so
much concerned with actual predictability but only
with determinism. In the framework of relativity we 
even can speak of a local determinism, i.e., the state of a
system at point $x$ and time $t$ is determined by
the configuration of the state at time $t-\Delta t$ (for
sufficiently small $\Delta t$, in order to avoid the
singularities mentioned in the footnote of the previous
paragraph) within a volume of radius $c \Delta t$. 
In classical physics we may choose $\Delta t$ to be as 
small as we like (keeping
it positive and non-zero). This local determinism 
(which is always present in the absence of an action
at a distance) extends to a global determinism (at
least within a finite part of our universe) and is
independent of the actual impossibility to make 
predictions due to the complexity of a system. 
For this reason this notion of determinism
extends also to non-linear systems with chaotic
behavior.

\subsection{The {\em tertium non datur} of predications}

In physics a predication about a system is a statement
about the values of observables which can be measured for 
this system.
In the following we will talk about pure states only.
Algebraically, a state can be defined as a positive,
normalized, linear functional on the set of observables.
These states form a convex set and pure states are the
states on the boundary of this set. On an operational
level a state can be defined by the equivalence class
of the history of the system,
where the history includes the preparation process.
Two histories are equivalent, if the 
corresponding states have identical expectation values
for all observables. For a pure state the history cannot
be refined in such a way that the variances of some 
observables become smaller without increasing the variances 
of other observables. 

In classical physics any observable has a well-defined
value for a pure state in the following sense: Each 
measurement of the same observable $F$ in the same 
state $\omega$ yields the same result $f$ which is
identical to the expectation value: $f=\omega(F)$.
There is no spreading of results, the variance of
the results for a sample of measurements of the same
observable is zero.
Therefore one can say, that in this situation the 
system has the property `$f$'. 
(There are technical problems when $f$ is a continuous 
variable, in which case one usually refers to 
intervals.) In this sense the {\em terium non datur}
- the exclusion of a third possibility - holds:
for a system in a pure state a predication is 
either true or false. 

In most cases we actually do observe variances of 
observables but in the context of classical physics 
it is taken for granted that these variances are 
either due to a mixture of states, i.e., the systems
under consideration are not prepared in a pure
state, or due to an experimental error, i.e.,
the uncertainties in the measuring procedure and
the limits in the precision of the measuring instruments
lead to a spreading of the data. A pure
system is assumed to have a definite value with
respect to any observable. 
   
\subsection{The separability of observer and observed}

A further assumption which is implicitly made in the
context of classical physics is the separability
of observer and observed. In the idealized case an 
observation (measurement) has no influence on the
observed system. In other words, the increase of
knowledge of the observer about a system does not
lead to a change of state of this system. 

At first sight this assumption seems to contradict
Newton's third law - ``actio'' equals ``reactio'':
the force of system 1 on system 2 equals
(in opposite direction) the force of system 2
on system 1. In a slightly different setting one
can also say that 
the energy loss of one system is equal
to the energy gain of the other system. 
Any observation obviously changes the state
of the observing system: the sensor of the measuring
device, the transformation to a change of the pointer 
of a measuring device up to the change of knowledge
on the side of the observer. Therefore, according to Newton's 
third law, this should be accompanied by a corresponding
change in the observed system. 

The more rigorous
statement of the classical assumption of a separation
between observer and observed is: The influence of the 
observing system on the observed system 
due to the observation (measurement) can
be made arbitrarily small and is independent of the
precision of the measurement. Equivalently, one
may say that in classical physics informational flow
does not need a corresponding flow of energy.
The changes on the side of the
observing system may be large and accompanied by 
a macroscopic amount of energy, however, 
this energy comes from an amplification mechanism.
A measuring device can be prepared in such
a way that an arbitrarily small change of one
component (a detecting device) is amplified in such
a way as to give rise to a macroscopically large 
change of another component (the pointer).

The independence of the observer from the
observed system should not be confused with
a ``God's-eye'' perspective on the side of the
observer, even though these two concepts are
closely related. A God's-eye perspective
(in contrast to an intrinsic perspective)
not only assumes that the observation is done
without disturbance of the observed system,
but it also assumes a preferred set of
measuring devices (clocks, rulers, etc.) which
are not subject to the standard laws of 
physics. Most prominent in the realm of
classical physics is the ``God's-eye'' perspective
in the context of relativity: clocks are
synchronized with respect to a prefered system,
the perception of events is not subject to
the delay due to the finite expansion
velocity of light, etc.  

\subsection{Sequential ordering of time}      

Newtonian mechanics (excluding
for a moment Maxwell's theory or the
theory of relativity) assumes a universal
time with a universal concept of simultaneity:
For any two events $a$ and $b$ in the universe
one of the following statements is true:
$a<b$, $a>b$ or $a=b$, where ``$<$'', 
``$>$'' and ``$=$'' refer to ``before'',
``after'' and ``simultaneous'', respectively.

In the context of special or general
relativity this is no longer true. We may
still define $a<b$ and $a>b$ as ``$a$ is in
the backward lightcone of $b$'' or ``$a$ is
in the future lightcone of $b$'', but there
is no universal notion of simultaneity. The most
one can say from an objective, universal level
is that two events $a$ and $b$ are causally
unrelated (i.e., they are in the causal 
complement of each other as defined by the
future and backward light-cones). In special 
relativity the notion of simultaneity is
used with respect to a given inertial system.
However, this requires a 
(to a large extend arbitrary) synchronization
convention for clocks within the same reference
system but at different positions. 

What still is true even in relativity
is the total sequential ordering of events
{\em along a world-line}.\footnote{At this
point we explicitly exclude non-causal solutions
of Einstein's equations which allow for
closed time-like loops like the Goedel
universe.} For any two events on a single
world-line the statements $a<b$, $a>b$ 
or $a=b$ are unambiguous. This is a
consequence of the necessary condition for
a world-line to be time-like. 
In this sense, time is
sequential along any world-line. (This statement
has to be modified for light-like world-lines 
corresponding to particles of zero
rest mass; in this case all events along 
this world line ``happen'' simultaneously.) 

We should remark that there is a corresponding
ordering structure for space. In 
Section \ref{sec_Categories}, the fourth slot
of our template of categories refered to
the ordering structure of space and time. 
However, in most cases we will only talk about
the ordering structure of time (the sequentiality
of time). The reason is that the topological ordering 
structure of space (giving meaning to ``inside'' and
``outside'' with respect to closed 2-spaces etc.) and
even more its metric structure (giving rise to relative
distances) is closely related to the localization 
of objects. The position of an object is generally 
considered as a property and, therefore,
can be treated in the context of the {\em terium
non datur}. The difference is due to an
asymmetry in physics with respect to the nature
of space and time: while we attribute
observables to the location or
position of an object, we usually do not 
introduce a ``time-observable''.   

\section{The categorical apparatus of the
statu nascendi aspect of reality}
\label{sec_E}

In the previous section we have briefly summarized
the categories (conceptual assumptions) applied 
in the context of classical (non-quantum) physics. 
None of these categories holds anymore in the
context of quantum theory. We now introduce the
corresponding categorical apparatus of quantum theory.

The four slots of our categorical template will be
filled with the following categories:
\begin{enumerate}
\item
Nature of processes: autogenetic.
\item
Combination of predications: paratactic.
\item
Relation between the observer and observed: 
self-referential. 
\item
Temporal ordering structure: 
time-space of the present.
\end{enumerate}
At first sight, these concepts may sound
unfamiliar, strange, and even undefinable. However,
we will show that essentially all of these concepts  
are already contained in the standard mathematical 
formalism used in quantum theory. However, we will use
these more general terms in order to emphasize that 
many of these concepts can also be found outside 
the realm of quantum theory.
(For a more philosophical account of these concepts
see \cite{AvM} and \cite{AvM2}.)

When we refer to quantum physical phenomena, usually
all four components of the second categorical apparatus
apply. However, we
see a particular close relationship between the 
following concepts of standard quantum theory 
with the expressions above:
\begin{enumerate}
\item
Autogenesis - the non-deterministic state reduction.
\item
Parataxis - the superposition principle.
\item
Self-referentiality - entanglement.
\item
Time-space of the present - the loss of sequentiality
for events.
\end{enumerate}
In the following sections we will make these 
concepts as well as some of the relations between
them more transparent.

\subsection{Autogenesis - The non-determinism in state reduction}

The classical Copenhagen interpretation of quantum
theory includes two processes by which the state of a 
quantum system can change in time: (1) the deterministic 
evolution of a closed quantum system according to 
Schr\"odinger's equation, and (2) the non-deterministic state 
reduction of a quantum system as the result of a measurement. 

While the first process of temporal change is largely 
undisputed, the second one is subject of ongoing debates. 
In particular protagonists of the many-worlds interpretation
(see Everett \cite{Everett} and deWitt \cite{DeWitt}) 
deny the existence of an ontic collapse even though
they usually ascribe an ontic reality to the wave function
(in contrast to information based interpretations of
quantum theory for which the quantum state itself has
only an epistemic meaning).
The fact that we seem to experience
a non-deterministic reduction of the quantum state is
explained by a rapid decoherence which makes it
impossible to construct observables which are able to 
interpolate between sufficiently different branches of 
the wave function and, therefore, it becomes increasingly 
impossible to observe interferences between these branches. 

As mentioned before, we will assume that the reduction
of a quantum state is related to an ontic part of our reality.
Presumably this is the strongest assumption about the
interpretation of quantum theory which we make in this 
context. According to this assumption, quantum theory is
intrinsically non-deterministic. This non-determinism
is not the consequence of a lack of knowledge, and
in this sense Leibniz' principle of sufficient reason
is violated in quantum theory.    

The relation of the non-determinism of quantum theory
with the ``status nascendi'' of quantum theory becomes
obvious when we notice that it is exactly the reduction
process which marks the transition from possibilities
to facts. The reduction process corresponds to a genuine
event and the results of this event are the facts which 
we ultimately observe.

Why did we name this category ``autogenesis''? 
In its original meaning, autogenesis means
``self generation''. By using this expression we want
to emphasize that the results of certain processes
are not predetermined by any external or internal cause. 
In the reduction process, one of several possibilities becomes
a fact and there is no cause whatsoever, which among these
possibilities will be realised. We should emphasize 
that ``autogenesis'' also excludes any internal cause in 
the sense of hidden variables within the system. It is 
the event itself, not some predetermined structure 
inside or outside the system, which leads to a particular 
outcome. 

\subsection{Parataxis - The Superposition Principle}

The second slot of our categorical template refers to
predications about physical attributes and, in particular,
how these predications may be combined. As we have
mentioned before,
in the F-scheme predications follow the standard
form of binary ``either-or'' logic (true or false).
Formally, this corresponds to a boolean lattice. 
Now we argue that in the second apparatus, the E-scheme, 
even contradicting predications
can stand side by side in the form of ``as well as''. 
We will refer to this property of the predication 
space as ``parataxis'', and its realization 
in quantum theory is the superposition principle.
In the context of a predication calculus, this
catergory may be realized by non-boolean lattices (see,
e.g., \cite{Giuntini}).

The concept of superpositions is most easily formulated
when quantum states are expressed as (normalized) vectors 
in a Hilbert space. Often it is stated in the form that 
with any two vectors $|\psi_1\rangle$ and $|\psi_2\rangle$ 
also the (normalized) linear combination
\begin{equation}
     |\psi\rangle = \alpha |\psi_1\rangle + 
          \beta |\psi_2\rangle
    \hspace{1cm} {\rm with~~} |\alpha|^2+|\beta|^2=1 
\end{equation}
is a quantum state.\footnote{Superselection rules 
may put restrictions onto this rule 
but this shall not concern us here.} 
Another formulation of the same
principle is that a state $|\psi\rangle$ can be expanded
in terms of the eigenstates $|\psi_i\rangle$ of any
self-adjoint observable $A$:
\begin{equation}
\label{eq_expansion}
   |\psi\rangle = \sum_i \alpha_i |\psi_i \rangle \hspace{1cm}
     {\rm with~~} \sum_i|\alpha_i|^2=1  \, , 
\end{equation}
where
\begin{equation}
    A |\psi_i\rangle = a_i |\psi_i\rangle \, . 
\end{equation}

In this formulation the superposition principle seems
to depend on the representation of pure quantum states 
as vectors of a Hilbert space. However, 
strictly speaking, pure quantum states 
rather correspond to the one-dimensional
rays in a Hilbert space, or, in other representations,
to the set of one-dimensional projection operators
on a Hilbert space or to the (convex) boundary of the set
of normalized, positive density matrices. For these 
objects one cannot define a unique addition. (The
normalized addition of projection operators or 
density matrices leads to mixed states while the
superposition of pure states is again a pure state.) 

A word concerning our notation: $\psi$ often
refers to a quantum state independent of its mathematical
representation. If we want to emphasize that the state is a
positive, normalized, linear functional on the set of
observables, we write $\omega_\psi$, and $\omega_\psi(A)$ 
denotes the expectation value of the observable $A$ in
this state. When 
$\psi$ is represented by a normalized vector, we write 
$|\psi\rangle$, and the expectation value of the
observable $A$ is
\begin{equation}
 \omega_\psi(A) = \langle \psi |A| \psi \rangle \, . 
\end{equation} 
Finally, if we represent the state by a projection operator
onto the ray defined by the vector $|\psi\rangle$,
we write $P_\psi$, and the expectation value of an observable
$A$ is
\begin{equation}
    \omega_\psi(A) = {\rm tr} (P_\psi A)  \, . 
\end{equation}

The superposition principle of quantum theory
expresses a property which is independent of 
the representation of pure states. We define
a state $\omega$ to be {\em dispersion-free}
with respect to an observable $A$, if
\begin{equation}
  \omega(A^2) = \omega(A)^2 \,. 
\end{equation}
When this property holds, a measurement of $A$ always 
yields the same result $a=\omega(A)$. 
In the F-scheme of classical physics any pure
state is dispersion-free with respect to any
observable.

If a vector can be expanded according to eq.\
(\ref{eq_expansion}), the corresponding state is not
dispersion-free with respect to the operator $A$,
unless all the expansion coefficients $\alpha_i$  
are zero except one coefficiet (which is one), in which 
case $|\psi\rangle$ is itself an eigenstate of $A$. 
More general, let 
\begin{equation}
     A = \sum_i a_i P_i 
\end{equation}
be the spectral decomposition of the observable $A$,
then
\begin{equation}
    \omega(A) = \sum_i a_i \omega(P_i)
\end{equation}
and
\begin{equation}
   \omega(P_i) = |\alpha_i|^2  \, . 
\end{equation} 
These values can be obtained from a distribution 
measurement of $\{a_i\}$. The relative phases
between these coefficients require measurements of
complementary observables (observables which do not
commute with $A$). In quantum 
mechanics it is impossible to construct pure states 
which are dispersion-free with respect to all 
observables \cite{Neumann}.

Let $\psi$ be a state which is not dispersion-free 
with respect to an observable $A$. In this case,
the same measurement (of observable $A$) performed on
the same state $\psi$ may sometimes yield the result
$a_1$ and sometimes a different result $a_2$,
where $a_1$ and $a_2$ are eigenvalues of $A$.
(For simplicity, we consider only
two possible outcomes.) In this case, we can 
neither say that
the state has the property $a_1$, nor
that is does not have this property. In this sense
the {\em tertium non datur} does not hold. Some
scientist prefer to say that this state $\psi$ is not
compatible with properties related to the observable $A$,
others will say that it has property $a_1$ 
``as well as'' property $a_2$. Fact is
that measurements of observable $A$ for the same
state $\psi$ sometimes yield the result $a_1$ and
sometimes the result $a_2$, and in this sense
both properties can be attributed to the state $\psi$. 

Parataxis is an expression which is used in philosophy
to denote that predications ``stand side by side'' and
that the {\em tertium on datur} does not hold. 
It is also used in the science of literatur where it refers
to a text in which a situation, phenomenon or object
is described by a
collection of (sometimes even contradictory) attributes.
Apart from the violation of the {\em tertium non datur}
two more properties characterize paratactic predication:
\begin{enumerate}
\item
The overall meaning of a paractic predication unfolds 
itself out of the constellations of components. This is
reflected in the mathematical formalism by the fact that
the transition amplitudes $\alpha_i=\langle \psi_i|\psi\rangle$
between the state $\psi$ and the states $\psi_i$, which 
correspond to the paratactic predicates $\{a_i\}$,
determine $\psi$ uniquely.
\item 
Formal conclusions are not possible and thus the
{\em ex falso quot libet} catastrophy (from one
contradiction one can derive any statement) is
avoided. Almost symbolic for this aspect in quantum
theory is the famous citation of Richard Feynman  
referring to the double slit experiment: ``{\em ... 
[The electron] always is going through one hole
or the other - when you look. But
when you have no apparatus to determine through which
hole the thing goes, then you cannot say that it either
goes through one hole or the other. (You can always {\em say}
it - provided you stop thinking immediately and make no
deductions from it. Physicists prefer not to say it, rather
than to stop thinking at the moment.) To conclude that it
goes either through one hole or the other when you are not
looking is to produce an error in prediction. ...}''
(\cite{Feynman} p.144).
Given the absence of formal truth criteria,
actual experience is the only way to verify a
proposition.   
\end{enumerate}

As we have seen, in classical physics either $a$ or
$\neg a$ is true, but in quantum theory $a$ as well as
$\neg a$ may be true in the sense, that we find 
states $|\psi\rangle$ which are a linear superposition
of an eigenstate $|\psi_a\rangle$ with the (dispersion-free) 
property $a$ and an eigenstate $|\psi_{\neg a}\rangle$ with 
the (dispersion-free) property $\neg a$:
\begin{equation}
  |\psi\rangle = \alpha |\psi_a\rangle
            + \beta |\psi_{\neg a}\rangle
   \hspace{1cm} {\rm with}~~~~
    |\alpha|^2+|\beta|^2=1 \, . 
\end{equation}
This property of quantum states is what we will 
denote by ``paratactic'', and $\alpha$ and $\beta$
are manifestations of the relative constellation of 
the components. 

\subsection{Self-Referentiality - the nonseparability of 
observer and observed}

One of the most significant and in it's consequences
most dramatic features of quantum theory is the
phenomenon of entanglement. It has to be kept in mind,
however, that entanglement 
can only be defined with respect to a tensor product 
representation of a Hilbert space ${\cal H}$:
\begin{equation}
   {\cal H} = {\cal H}_1 \otimes {\cal H}_2 \, . 
\end{equation} 
A state $|\Psi\rangle$ (represented by a normalized vector)
in ${\cal H}$ is called {\em separable} if it can
be written in the form 
\begin{equation}   |\Psi\rangle 
  = |\psi \rangle \otimes |\varphi\rangle  \, , 
\end{equation}
where $|\psi\rangle \in {\cal H}_1$ and 
$|\varphi\rangle \in {\cal H}_2$.
A state which is not separable is called 
{\em entangled} (with respect to the tensor
product representation).

Typical entangled states are {\em EPR-states}
\begin{equation}
   |\Phi_{\rm EPR}\rangle = \frac{1}{\sqrt{2}} \left(
   |\psi_1\rangle \otimes |\varphi_2\rangle -
   |\psi_2 \rangle \otimes |\varphi_1\rangle \right) \, , 
\end{equation}
or, more general, {\em Bell states}:
\begin{equation}
   | \Phi_{\rm Bell}\rangle = \frac{1}{\sqrt{2}} \left(
   |\psi_1\rangle \otimes |\varphi_2\rangle  \pm
   |\psi_2 \rangle \otimes |\varphi_1\rangle \right)  
\end{equation}
and
\begin{equation}
   |\Phi'_{\rm Bell}\rangle = \frac{1}{\sqrt{2}} \left(
   |\psi_1\rangle \otimes |\varphi_2\rangle \pm
   |\psi_2 \rangle \otimes |\varphi_1\rangle \right) \, . 
\end{equation}
With respect to measures of entanglement these
states are maximally entangled. 

In a similar way entanglement can also be 
defined for higher tensor product representations 
of the state space. For these higher order tensor products
one can even define different types of entanglement.
A famous example are the GHZ-states for three particle
systems \cite{GHZ}.

In the context of quantum theory, the tensor product 
representation is interpreted as a splitting or
a partition of the
system described by ${\cal H}$ into two subsystems
which are described by ${\cal H}_1$ and ${\cal H}_2$,
respectively. In many cases, such a partition seems
natural, e.g., when the total system consists
of two particles and the subsystems refer to the
single particles. In other cases such a splitting
may appear arbitrary, e.g., when the description of 
a single particle is split with respect to the 
coordinates of its reference system. 
Another, more relevant, example is a single particle 
for which the total state space
is splitted with respect to the spatial degrees of 
freedom on the one hand and the spinorial degrees
of freedom on the other hand. For an electron which passed
a Stern-Gerlach magnet one may speak of an entanglement
between the spatial position of this electron and
its spin orientation:
\begin{equation}
   \Psi(x,s) = \frac{1}{\sqrt{2}} \left(
   \psi^+(x) \otimes |s=+\rangle +
   \psi^-(x) \otimes |s=-\rangle \right) \, . 
\end{equation}
$\psi^{\pm}(x)$ are parts of the spatial wave function
which are peaked around different positions. 

The weird aspects of entanglement enter when the
system described by the quantum state is non-local, e.g., 
when the system consists of two particles which are 
far apart. A measurement performed at one particle
has immediate consequences for the state of the other
particle. To be more precise, a measurement performed
at one particle leads to a reduction of the total
state in such a way that the state becomes separable,
which in turn makes it possible to assign a definite state 
to the second particle. According to the standard 
formalism of quantum theory, this state reduction occurs
instantaneously. As the resulting correlations do not
allow to transmit information or energy, this
process does not violate the causality principle
of the theory of relativity. Nevertheless, Einstein
called this ``superluminal'' change
of the state due to the reduction process 
a ``spooky action at a distance''. (It should be
mentioned that this ``spooky action at a distance'' not
only occurs for entangled systems but also for single 
particle states with a highly non-local wave function.)

In principle, any interaction between two particles
- or, more generally, between two systems - leads to
entanglement. (On the other hand, subsequent interactions 
with other systems may destroy these entanglements.)
Up to now, we mostly considered the interaction between
particles or typical quantum systems. However, according
to the standard formalism of quantum theory, this statement 
remains true for macroscopic systems. A particular case
is the situation of a measurement, where a quantum
system (a system with few degrees of freedom) interacts
with a classical system (a system with many degrees of
freedom). In quantum theory the measurement process
has to be considered as an interaction which has an
influence on both systems, the observing as well as the 
observed system. As a result of this interaction the
observed system and the observing system become
entangled. We briefly recall this situation in the
usual framework.

Let 
\begin{equation}
  |s\rangle = \sum_i \alpha_i |s_i\rangle 
\end{equation}
be the state of a quantum system, expressed as
a superposition of eigenstates with respect to
the measured observable. (We are not addressing
the ``pointer basis problem'' here, 
see e.g.\ \cite{Zurek}, so we assume
that these states are uniquely given.) Let
$|\varphi_0\rangle$ be the initial state of the
measuring device, and $\{|\varphi_i\rangle\}$
be its pointer basis. The initial state of
the combined system $|\Phi_0\rangle$ is separable:  
\begin{equation}
\label{eq_measurement1}
  |\Phi_0\rangle = |s\rangle |\varphi_0\rangle =
    \sum_i \alpha_i |s_i\rangle |\varphi_0\rangle \, . 
\end{equation}
The interaction between the two systems due to the 
measuring process leads to a superposition of
correlated states:
\begin{equation}
\label{eq_measurement2}
 |\Phi_1\rangle = 
  \sum_i \alpha_i |s_i\rangle |\varphi_i\rangle \, . 
\end{equation}
This state is an entangled state. The observing system
and the observed system are no longer separated. It 
is impossible to assign a separate quantum state to 
any of these two systems, and only the quantum state
corresponding to the combined system is meaningful.
In a next step the reduction process describes
the transition from the superposition of
possibilities as expressed in 
eq.\ (\ref{eq_measurement2}) to a single fact:
\begin{equation}
\label{eq_measurement3}
  |\Phi_1\rangle ~\longrightarrow ~
  |\Phi_2\rangle = |s_k\rangle |\varphi_k\rangle \, .
\end{equation}

Hence, the process of observation first leads
to entanglement and, therefore, an
inseparability between observer and observed.
In a second step the reduction process results in
physical facts. Obviously also the other two 
concepts discussed so far play an essential role: 
the paratactic predication (expressed in the 
superposition of possibilities) and the 
intrinsic non-determinism of the reduction.

We have attributed the loss of dichotomy between
observer and observed to the general concept of
self-referentiality. The philosophical reasons
for this attribution will be discussed elsewhere
\cite{AvM2}. However, it should be obvious
that when observer and observed are no longer
separated, we have the situation that ``a system
observes itself'', which is a strong form of
self-referentiality.

\subsection{The non-sequentiality of time}

The final slot of the categorical template refers to
the ordering structures of space and time. As we have
mentioned before, the loss of the relational metrical 
ordering in space (as defined, e.g., by relative distances) 
as well as the topological ordering (e.g., the 
notion of ``inside'' and ``outside'' with respect to
closed 2-surfaces) is already obvious for systems with
spatially extended wave functions. It can be associated
with a loss of ``either-or'' predications with respect to 
the position.
Therefore, we will concentrate in this section on the 
partial loss of the predominant ordering principle of 
time - the sequentiallity of events along a world line - 
in quantum theory. 
Of course, this loss of sequentiality is
restricted to typical quantum phenomena, i.e.\
in most cases to very short time intervals. Only in
special cases are the relevant time intervals 
of macroscopic extension.

We first have to specify what is meant by ``loss
of sequential ordering''. In quantum theory, time is
treated as a classical 1-parameter variable which, 
of course, allows for a well-defined ordering. This
variable refers to a mathematical time and it is
realized by a
classical clock (we will say more about this point
later). However, as we have described in Sec.\ \ref{sec_F}, 
the sequentiality of time in the F-scheme refers to the
sequentiality of events along world-lines: for any
two events $a$ and $b$ on a world-line one can say
``$a$ before $b$'' or ``$b$ before $b$''. It is this 
property which is partially lost in quantum theory.   

Let us consider the quantum propagator of a system,
i.e., the probability amplitude $K(x,y,t)$, where $x$ 
is the initial state of the system, $y$ the final state,
and $t$ the propagation time (as measured by a classical
clock). Here $x$ and $y$ may represent the position of a single
particle, the collective positions of many particles,
the configuration of a field etc. According to Feynman's 
representation of this propagator as a sum over histories
we may write:
\begin{equation}
   K(x,y,t)= \sum_{x\rightarrow y} \exp \left(
     \frac{{\rm i}}{\hbar} S[x\rightarrow y;t] \right) \, . 
\end{equation}
``$x\rightarrow y$'' represents a possible history
of how the system can evolve from state $x$ to state $y$
within the time interval $t$ (which is the same for all
histories); $S[x\rightarrow y,t]$ is the classical action
for this history, and $\sum_{x\rightarrow y}$ symbolizes 
the summation over all such histories. If these histories 
involve certain events $a_1$, $a_2$, ... , and if the
general constraints on the set of histories do not forbid
a permutation of the temporal order (with respect to the
mathematical background time $t$) of these events, then
the summation over histories implies also a summation over 
all these permutations. In such a case, when a system
has propagated from state $x$ to $y$, it is meaningless 
to state that a certain event $a_1$ happened before a second
event $a_2$ or vice versa. The classical sequentiality
of events along the world line of the system is
lost. 

The loss of sequentiality as it follows from the summation
over histories representation is most obvious in
elementary particle physics, where the amplitudes for
scattering processes are expressed in terms of Feynman
digramms. This representation includes a summation over 
all possible types of events as well as an integration 
over all possible space-time locations of these events. 
In this form not only the loss of sequentiality is obvious,
but also the loss of determinism, the paratactic 
predication (all possible processes actually contribute)
and even the self-referentiality in the sense of a
highly interactive and entangled system.

Two objections may come into mind at this point: (1) In the
examples given above one cannot even say {\em that} a certain
event has happened, therefore, the loss of temporal sequentiality
may rather be due to a loss of facticity. 
(2) Is the loss of temporal
sequentiality not an immediate consequence of the energy-time
uncertainty relation in quantum theory?
Both issues shall be addressed now.

(1) Let us consider an example where it is known that two different
events $a$ and $b$ have happened, but where the temporal order
of these events is open. Imagine two different (distinguishable) 
atoms $A$ and $B$ in a small box. Both atoms shall be in an
exited state, and the transition energy from the excited state
to the ground state shall be equal in both
cases. We will observe two photons of the same energy
emerging from the box, possibly with a large temporal delay
between the first and second event. However, without disturbing
the system by an additional observation, we cannot tell which
of the two observed photons corresponds to the decay
of which atom. In fact, in the summation over histories we
have to sum over both possibilities (atom $A$ decaying first
and $B$ second and vice versa), and again the temporal order
of these events (which are now known to have happened) is
not defined. This situation resembles the spatial
non-separability in the double slit experiment: It
is known that a particle has passed through a 
double slit but there is no information about which of
the two slits it has passed through.

(2) The uncertainty relation between time and energy,
$\Delta t \cdot \Delta E \geq \hbar/2$, refers to a 
limitation in the precision of measurements of these two 
quantities on the same system: $\Delta E$ is the 
uncertainty in an energy measurement during a time interval 
$\Delta t$, or, vice versa, $\Delta t$ is the maximal 
precission for the determination of the moment of an 
event in which an energy is transfered of which the value 
can be known up to $\Delta E$. The example given above 
(the decay of two atoms) involves the energy of the 
emitted photons. For each single atom 
the uncertainty of the energy
of the emitted photon is related to the precision with
which the moment of the decay can be determined. If both
decays happen within these time ranges, the loss of
sequentiality can be attributed to the uncertainty
relation between time and energy. (For similar examples
related to the loss of sequentiality see, e.g., 
\cite{Aharonov,Oppenheim,Oppenheim1}.)

However, concerning the sequentiality of the two decays, 
what is relevant is the indistinguishability of the emitted
photons. If the photons have the same energy, it is
impossible to attribute one of the photons to one of the 
events and the sequence of events cannot be determined.
If, however, for some reason, atom $A$ only emitts photons
with a left circular polarisation and atom $B$ only emitts
photons with a right circular polarisation, the association
of an emitted photon to a certain event $a$ or $b$ would
be possible and the sequentializability indeed only depends
on the precision of how exact the moments of emission
can be measured, which again would be subject to the 
energy-time uncertainty relation. The two cases - polarized
or unpolarized photons - involve the same energies, however,
the loss of sequentiality only refers to the unpolarized 
case.

\section{The time-space of the present}
\label{sec_Time}

In the previous section we described the loss of
sequentiality of events within certain time
limits. In this section we will draw some conclusions
from this observation and, in particular, emphasize
that the quantum mechanical formalism gives strong 
indications that a complete unified theory of space 
and time on the one side with the principles of 
quantum theory on the other side may not be
possible without a theory of the present.

The paratactic predication with respect to the
location of objects together with the loss
of temporal sequentiality of events make it
difficult if not impossible to obtain a ``block universe'' 
picture of reality. By ``block universe'' we mean a 
well-defined and fixed space-time representation of the 
universe. (Famous is the description of a block universe
by Sir Arthur Eddington: ``Events don't happen; events
are simply there'' \cite{Eddington}.)
The fact that the locations of particles
and the sequentiality of events are not determined
within certain limits leaves ``bubbles'' in a
block universe description for which, due to the 
inherent inseparability, a higher resolution
description is not possible.
Similar bubbles occur also in the ``consistent
histories'' descriptions of quantum theory 
\cite{Griffiths,Omnes}
where it becomes obvious that a refinement of
the set of possible histories to one single history
describing the facts in our universe is impossible. 

Of course, such bubbles only remain when one
tries to represent the histories in terms of
classicle particle trajectories. If, instead, one
considers the wave function (or, equivalently,
the quantum state) as the essential entity,
a block universe representation is possible, but 
due to the indeterminacy of quantum theory
and the reduction process, certain branches of
the wave function may come to an end while
other branches become enhanced. In both cases
a prediction of the future state of the
universe in terms of a present state is impossible.
Within the many-worlds interpretation
of Everett \cite{Everett,DeWitt} this problem
is circumvented, because there
is no collapse of the quantum state and the
block universe representation in terms
of quantum states is deterministic.

However, one problem remains for all of
these representations. The standard formalism
of quantum theory seems to indicate the
existence of a prefered simultaneity - quantum
theory is non-local. Consider an EPR-state
where the two entangled particles are far
apart. A measurement on one of the particles 
leads to an instantaneous non-local reduction of 
the quantum state (in the
many-worlds representation, the wave function 
splits non-locally into two branches).
If we assign an objective (ontic) meaning to
a quantum state, this non-local reduction
would be a non-local event which distinguishes
a prefered simultaneity at two distant locations.
One could even imagine an entangled state consisting 
of many particles densely distributed over the whole
universe,
\begin{equation}
  |\Psi\rangle = \frac{1}{\sqrt{2}} \left(
  |+,+,+,+,...\rangle + 
  |-,-,-,-,...\rangle \right) \, , 
\end{equation}
where ``$+$'' and ``$-$'' refers to some
spin orientation of the single particles.
A measurement of one particle leads to the
reduction of the state of all particles and,
therefore, to a global notion of simultaneity. 

Among the standard
interpretations of quantum theory, only a
subjective interpretation of the quantum state
- describing just the information we have 
about a system - 
circumvents this problem, because in this
interpretation the reduction of the quantum state 
has no observer independent ontic 
correlate.\footnote{In the GRW-formalism
\cite{GRW} a Lorentz-invariant reduction 
process has been developed. However, 
many of the concepts employed in this
approach - a multi-time formalism, non-local
correlations in the location of reduction
points, and a stochastic (non-deterministic)
extension of the Schr\"odinger equation -
indicate a close relationship to our
E-scheme categories.} 

Our formalism circumvents the problems 
associated with non-local reduction processes.
The locality requirement and the causal structure
of relativity refer to the factual aspects
of reality. When a measurement is made on
one of the particles, this becomes a fact
in the vicinity of this particle. Even though
the state is reduced simultaneously for the second
particle, as long as no measurement is made 
at the second particle, this reduction is not
factual in the vicinity of this particle. 
(Of course, what is relevant here is not the
measurement as an experimental situation, but
the question whether the second particle
has sufficient interaction with its environment
in order to transfer the information about the
reduced state to its environment; in this case
this information would become factual.)
On the other hand, if a measurement is made at the
second particle, the result coincides with the
first measurement. However, as the experimenter 
has no influence on the result of the measurement,
it is impossible to transfer a signal using
the reduction process.

Quantum theory gives us at least two indications
that for a full understanding a theory of the
present might be necessary: (1) the non-deterministic
reduction process describes the transition from
potentialities to facts, which is exactly what
one would expect to occur in a present, (2) the
non-local simultaneity distinguishes a global
space-like hyperspace where reduction processes
occur simultaneously, and this would be a 
prerequisite for
an ontic (global) ``present''. It has been argued
at several occasions that a theory of the
present contradicts special and general relativity.
Einstein himself, in a discussion with R.\ Carnape,
was disappointed that the theory of relativtiy does
not allow for a theory of the present \cite{Einstein_Carnap}, 
and a famous argument in this discussion is due to K.\ 
Goedel \cite{Goedel}. (For Goedel it was not the
Minkowski space itself which contradicted a theory
of a present, because a distinguished space-like 
hyperspace is at least possible, but due to the theoretical
existence of solutions of Einstein's equations with
light-like closed loops - as, for instance, in 
the Goedel universe - he considered a theory of the 
present as impossible.)

We want to indicate now some of the ingrediants of
such a theory of the present. In accordance with the
previous section, the present has an extension.
This extension is attributed to events, it depends 
on the process (i.e.,
it is not the same for all processes), and it is
measured against an external clock. These statements
need a clarification which will be given now. 

The extension of the present is indicated by the
transition from virtual possibilities to facts.
This transition is not sharp. Due to certain 
properties, a particle may interact with its 
environment and, thereby, transfers the information
about these properties to environmental degrees of 
freedom which now, due to their interaction with
ever more degrees of freedom, spread
this information to an increasing number of other
particles or systems. By this process, events become
facts. Facticity is not an either-or property, but
it increases gradually. It is closely related to the
concept of decoherence. A (qualitative) measure of 
facticity would be the effort to ``undo'' a certain 
event, i.e., to restore all the coherence necessary 
in order to generate interference effects which
indicate that a certain event did not happen. 
(It has been estimated \cite{Omnes} that in order
to ``undo'' $n$ degrees of freedom which took part in
decoherence, a system of the order of $\exp(a n^{3/2})$
degrees of freedom is necessary, where $a$ is some
positive constant; this indicates a possibility to 
quantify such a measure.) Therefore, facticity
is never absolute but only FAPP (``for all practical 
purposes'', an expression introduced by J.\ Bell
in a similar context \cite{Bell_against}).

In some processes the extension of the present
is very short (less than nano-seconds), like, e.g.,
when a particle his a screen where it is registered.
Under special experimental conditions,
one can prolong the process of factualization
almost arbitrarily long (e.g.\ in so-called
quantum eraser set-ups \cite{Scully}). An extreme
example would be the scenario of the cosmic
delayed choice experiment which is attributed to 
Wheeler\cite{Wheeler}. Depending on the experimental
set-up on earth one can decide to measure whether
a photon emitted from a quasar several billion
years ago passed by a gravitational lense on
both sides (in a double-slit manner) or on
one side. Concerning the history of this photon
the present extends over billions of years.

Even within a single process, the extension of the
present attributed to different events may differ.
One can rapidly measure the energy of an electron
in an electric field without determining its spin
orientation. For atoms one can measure the
absolute values of the magnetic moments in an 
electric field without determining their sign. Therefore,
facticity (and hence also the extension of a present)
has to be attributed to events and not to systems. 

By an external clock (against which we measure the
extension of the present) we mean a system for which
the time intervals of the transitions to facticity 
are extremely short.
Today, atomic clocks or fountain clocks
can resolve time intervals which are much shorter
than the ``extension of the present'' for many of
the systems mentioned above. Presumably there is
a principal lower bound for the resolution of such
``external clocks'' (which might be the Planck
scale of roughly $10^{-44}$\,s), which would be a 
lower bound for the extension of the present of
any event.  

\begin{figure}[htb]
\begin{picture}(400,100)(0,0)
\put(10,10){\vector(1,0){150}}
\put(10,10){\vector(0,1){80}}
\put(80,8){\line(0,1){4}}
\thicklines
\qbezier(10,10)(70,10)(80,50)
\qbezier(80,50)(90,90)(160,90)
\thinlines
\put(210,10){\vector(1,0){150}}
\put(210,10){\vector(0,1){80}}
\put(280,8){\line(0,1){4}}
\thicklines
\qbezier(210,10)(255,10)(265,40)
\qbezier(265,40)(272,80)(280,80)
\qbezier(280,80)(288,80)(295,40)
\qbezier(295,40)(305,10)(350,10)
\put(165,5){\makebox(0,0){$t$}}
\put(80,4){\makebox(0,0){$t_0$}}
\put(25,95){\makebox(0,0){\footnotesize Facticity}}
\put(225,95){\makebox(0,0){\footnotesize Measure of Present}}
\put(280,4){\makebox(0,0){$t_0$}}
\put(365,5){\makebox(0,0){$t$}}
\end{picture}
\caption{\label{fig_present}%
(left) Facticity (as measured by the effort to
undo an event, see text) against an idealized external
time. (right) The derivative of facticity yields a measure
for the intensity of the present. The time $t$ refers
to an external classical clock.}
\end{figure}
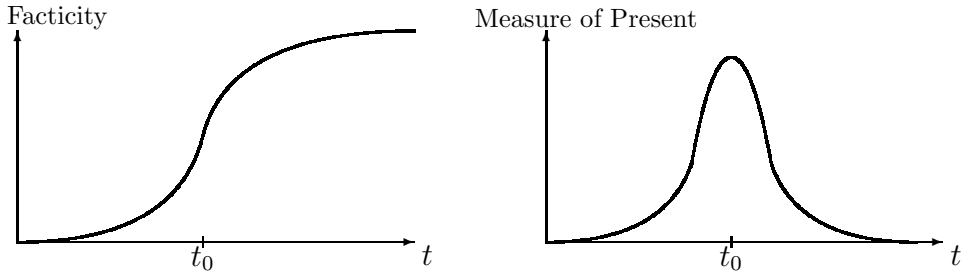

We thus arrive at the following picture (see Fig.\
\ref{fig_present}): For each event there is a
measure of facticity whose derivative (with 
respect to the external ``classcial'' time) 
gives rise to a measure for the intensitiy of the 
present of this event. Entanglement leads to 
a correlation of such curves for different
processes and at different locations thus giving 
rise in a classical limit to a globalized meaning 
of the present. On the level of quantum processes, 
this meaning is never absolutely precise.

\section{Summary and conclusion}

In this arcticle we hoped to achieve two goals:
(1) We have clarified the conceptual foundations
of quantum theory and put them in opposition to
the conceptual foundations of classical physics.
(2) We have outlined certain aspects of a
``theory of the present'', which in our opinion
is inherently already present in the existing
formalism of quantum theory.

The E-scheme of the conceptual foundations
of quantum theory is complementary to the
F-scheme of the conceptual foundations of
classical physics. Together they constitute
the conceptual framework which is 
necessary to make statements about reality.
In the present article we have elaborated the
E-scheme only in the context of quantum theory.
However, to a certain degree the E-scheme
is also relevant for other areas of science.
The ingredients - non-determinism, 
self-referentiality, non-separability between
observer and observed, etc. - can also be found
in other (complex) systems. In \cite{FM}
we have argued that both schemes are relevant
in addressing the problem of conciousness.
In other areas - e.g.\ in certain evolutionary
processes or in addressing the
question of ``What is life?'' - it may also
turn out to be relevant that both families
of concepts are taken into account.

The concept of a ``time-space of the present''
which we developed in the last part of this
article, should indicate a new approach 
towards a theory of the present. We believe
that an understanding of the fundamental concepts
of nature forces us to
develope such a theory, and quantum theory
already gives us strong hints towards a
non-subjective existence of the present. 
Next to consciousness, the present is one of 
our most intensive experiences, and without
a theory of both we will always lack an 
understanding of the most
profound basis of reality. 
\vspace{1cm} 

\noindent
{\large\bf Acknowledgements}
\vspace{0.3cm}

\noindent
It is a pleasure to acknowledge numerous
discussions with friends and collegues about
the ideas presented in this article. A list can
never be complete, but we are particularly grateful
for interesting suggestions from Harald Atmanspacher, 
Hans-Peter D\"urr, Ernst P\"oppel, Michael Silberstein, 
and E\"ors Szathmary, as well as from the participants 
of a Parmenides-Workshop on ``Philosophical and
Physical Perspectives of Time and the Present'' held
in October 2006, in particular Domenico Giulini, 
Michael Drieschner, Erich Joos, Claus Kiefer, and 
Tejinder P. Singh.

\end{document}